# Elongated Nano Domains and Molecular Intermixing induced Doping in Organic Photovoltaic Active Layers with Electric Field Treatment

Rabindra Dulal[1], Akshay Iyer[2], Umar Farooq Ghumman[2], Joydeep Munshi[3], Aaron Wang[1], Ganesh Balasubramanian[3], Wei Chen[2], and TeYu Chien[1]

[1]Department of Physics & Astronomy, University of Wyoming, Laramie, Wyoming 82071, USA

[2]Department of Mechanical Engineering, Northwestern University, Evanston, IL 60208, USA

[3]Department of Mechanical Engineering & Mechanics, Lehigh University, Bethlehem, PA 18015, USA

**Abstract**

The effects of the electric-field-assisted annealing on the bulk heterojunction nano-morphology in the P3HT/PCBM active layer of the organic photovoltaic cells (OPVCs) are presented here. It was widely accepted that the electric-field-assisted annealing will facilitate the P3HT, the polar polymer, to be better crystalline to enhance the charge mobility, hence the improvement of the OPVC performance. The influences on the nano-morphology of the electron donor and accepter domains are not well understood. Here, using the cross-sectional scanning tunneling microscopy and spectroscopy (XSTM/S), the electric-field-assisted annealing treatment is found to influence the molecular domains to be elongated with the orientation near the direction of the external electric field. The elongation of the molecular domains is believed to facilitate the domain percolation, which causes higher charge mobility, hence the higher short-circuit current density ($J_{sc}$). On the other hand, it was also observed that the electronic properties of the P3HT-rich and PCBM-rich domains in the electric-field-assisted annealed samples showed smaller energy band gaps and smaller molecular orbital offset between the two domains, which is argued to decrease the open circuit voltage ($V_{oc}$) and negatively impact the OPVC performance. Based on the X-ray diffraction (XRD) and small angle X-ray scattering (SAXS) results, the altered electronic properties are argued to be due to the molecular intermixing induced doping effects. These results point out competing factors affecting the OPVC performance with the electric-field-assisted annealing treatment.

**Introduction:**

With the properties of environment friendly, flexibility and low cost, organic photovoltaic cells (OPVC) are considered as one of the most promising next generation photovoltaic technologies [1]. P3HT (poly-(3-hexyl-thiophene)) and PCBM (phenyl-C61-butyric acid methyl ester) are one of the most studied organic semiconductor pairs for OPVC applications [2–4]. Bulk heterojunction (BHJ) architecture of OPVC active layer is believed to be responsible for the high efficiency due to its quasi random nanomorphology, which facilitates charge separation for the excitons with short exciton mean free path in the organic molecules ( in the order of 5 – 10 nm) [5]. Besides identifying the suitable organic molecule combinations [1], most efforts were focused on how to maximize power conversion efficiency (PCE) through varying synthesis parameters, which directly affect the nanomorphology of the electron donor-rich and acceptor-rich domains [6]. Among the various synthesis parameters [6–14], the effects of the application of electric field during synthesis or as a postproduction treatment have shown improvements in the PCE [11,15–22]. A better understanding of the electric field treatment will provide great insights on further improving the OPVC performance.

Various types of the electric field treatments have been reported in literature. The early research with DC electric field treatment was done at spin-coating stage or not at an elevated temperature (room temperature drying) [23–25]. It was pointed out that the DC electric field post-production treatment needs to be at elevated temperature to be effective [11,26,27]. It was argued that the enhanced OPVC performance is mainly due to the improved hole mobility (along the electric field direction) with the better crystalline polar polymer molecules [11,23,28–30]. Similar effect upon the application of the external electric field at elevated temperature was also reported for organic light-emitting diodes (OLEDs) [31]. Other factors were also discussed in literature, such as the elimination of the nano-domains (better intermolecular mixing) [32] or the rougher surfaces for better contact [24]. In some works, the electric field dependence experiments showed that there exists a certain optimized electric field strength for the optimal OPVC performance [20–22]. This observation indicates that the electric field treatment may induce both positive and negative impacts on the OPVC performance. So far, there is no fundamental understanding regarding the effects of the electric-field treatment on the OPVC performance. The main issue is the lack of proper tools capable of probing the subtle change in the electronic properties and the nanomorphology in the active layer with high spatial resolution. In particular, "*how does the electric-field-assisted annealing treatment affect the molecular intermixing and the nanomorphology of the donor- and acceptor-domain?*" is the key question to be answered here.

In this study, the effects of the electric-field-assisted annealing treatment on the nanomorphology and the molecular intermixing as well as the local electronic properties are studied by cross-sectional scanning tunneling microscopy and spectroscopy (XSTM/S). STM/S is capable of probing electronic local density of states (LDOS) near Fermi energy through $dI/dV$ measurements with spatial resolution easily down to nm scale or, sometimes, sub-nm scale. With strong difference in highest occupied molecular orbital (HOMO) and lowest unoccupied molecular orbital (LUMO) between electron donor and acceptors, the $dI/dV$ signals measured in donor-rich and acceptor-rich regions will differ significantly, giving the molecule sensitivity needed for OPVC active layer measurements [33–36]. On the contrary, atomic force microscopy (AFM) [15,37–39] can only be sensitive to the morphology without molecular sensitivity and the spatial resolution is limited by the size of the tip apex (~10-20 nm). The transmission electron microscopy (TEM) [18,40–42] has excellent spatial resolution but lacks molecular sensitivity, as pointed out by Kiel *et al.* [43]. The main reason to use the XSTM/S, instead of top-down STM measurement, is to prepare freshly fractured P3HT/PCBM surfaces in the cross-sectional view for STM measurements without further modifying the nanomorphology or molecular intermixing by further vacuum annealing (by typical sputtering/annealing preparation for top-down STM measurements) [35,36]. The results show that the electric-field-assisted annealing will induce: (a) elongated molecular domains aligned closely with the electric field direction; and (b) different levels of the molecular intermixing between PCBM and P3HT, which further affects the LDOS in the donor (P3HT-rich) and acceptor (PCBM-rich) domains. These results provide important insights on how does the electric-field-assisted annealing impact the OPVC performance.

**Sample Preparation and STM Measurements:**

P3HT (regio-regular (RR 93-95) SOL4106, used as received from Solaris Chem Inc.) and PCBM (purity >99.5%, SOL5061, used as received from Solaris Chem Inc.) were made into separate solutions in chlorobenzene (purity ≥99.5%, sigma-aldrich) with 1.78 wt % concentration. Solutions with desired P3HT:PCBM weight ratio 1:1 were then made by mixing the precursor solutions, followed by spin coating onto the Si(100) substrate with 1,050 rpm for 1 minute. The P3HT: PCBM/Si(100) films were annealed at 100 °C for 20 minutes in inert environment. Electric field ($3.89 \pm 0.04$ kV/cm) treatment was performed during the annealing process with all other synthesis parameters remained the same. In particular, the spin coated films were placed in between two metal plates with a gap of $0.90 \pm 0.01$ mm while $350 \pm 1$ V potential difference was applied on the two metal plates. The whole capacitor-like stage was placed on hot plate for simultaneous annealing process. Figure 1(a) illustrates the setup of the electric field application during the annealing process (electric-field-assisted annealing process).

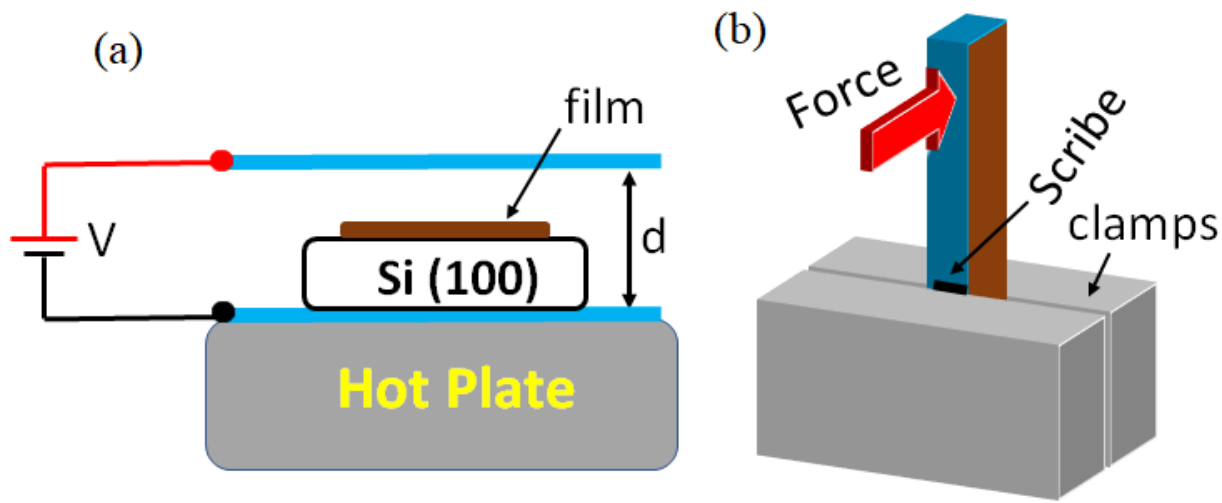

Figure 1. (**a**) Experimental setup for the electric-field-assisted annealing. (**b**) Experimental setup for the XSTM sample preparation.

The treated samples were mounted on a XSTM sample holder and fractured in ultra-high vacuum (UHV) environment prior to the XSTM/S measurements. XSTM/S has been used to image OPVC active layers [33,36] and perovskite solar cell materials [45]. Fracturing of the thin films generates fresh, cross-sectional surfaces for XSTM/S measurements [44]. This preparation method is suitable for OPVC active layer measurements as the typical sputtering/annealing method for top-down STM measurements will change the targeted nanomorphology in the OPVC active layer.

Figure 1(b) shows the sample fracturing geometry using a home-made XSTM sample holder for XSTM/S measurements with sample thickness of 0.5 mm, length of 8 mm and width of 1 mm. The sample fracturing was done by first scribing the side of the sample followed by the fracturing process from the side of the film, as illustrated in Fig. 1(b). This geometry will ensure high quality (sharp edge) films ready for XSTM/S measurements [45]. The sample fracturing was done in the UHV environment with the base pressure of $10^{-9}$ mbar and scanned under the pressure of $10^{-11}$ mbar. All the data were collected at room temperature.

**X-ray diffraction (XRD) measurement:**

The X-ray diffraction (XRD) and small angle X-ray scattering (SAXS) were measured on the films prepared (1) without annealing; (2) with annealing and (3) with electric-field-assisted annealing to gain insights on how the crystallinity and nanomorphology are influenced by the treatments. The measurements were done with a Rigaku Smartlab diffractometer with Cu target ($K_\alpha$ wavelength of 1.542 Å). The XRD measurements were performed from 3° to 14° ($2\theta$) with scanning step of 0.05°. The SAXS were measured from 0.0052° to 1.12° with scanning step of 0.0052°. For the SAXS, the $2\theta$ values were converted to corresponding $\Delta k$ value (denoted as $k$ hereafter for simplicity) using the relation, $\Delta k = \frac{2\pi}{\lambda} 2 sin\theta$, where $\lambda$ is the wavelength of the X-ray. With this conversion, the SAXS data (Fig. 4(b)) range is from 0.004 nm$^{-1}$ to 0.8 nm$^{-1}$.

**Result and Discussion**

Figure 2(b)-(e) show the topography and the d$I$/d$V$ mappings of two samples: (1) annealed without electric field treatment; and (2) with electric-field-assisted annealing. Figure 2(a) shows the height profiles along the black line in Fig. 2(b) and the red line Fig. 2(d), respectively. The roughness for the electric-field-assisted annealed sample is found to be ~1.2 nm while that for the samples without electric field treatment is determined to be ~0.2 nm. Similar trend was reported on the top surfaces measured by AFM [15,16,24]. For both samples, the topographies (Fig. 2(b) and (d)) exhibit anisotropic features, which are likely due to the fracturing process [45]. This anisotropic topography does not necessarily indicate that the molecular domain textures are anisotropic. The molecule-sensitive d$I$/d$V$ mappings (Fig. 2(c) and (e)) show distinct textures between the two. While an isotropic random distributed two domains are observed in the without electric field treated sample (Fig. 2(c)); anisotropic and elongated molecular domains are observed in the electric-field-assisted annealed samples (Fig. 2(e)). It is already clear here that the electric-field-assisted annealing treatment for the P3HT/PCBM active layer induces the elongated

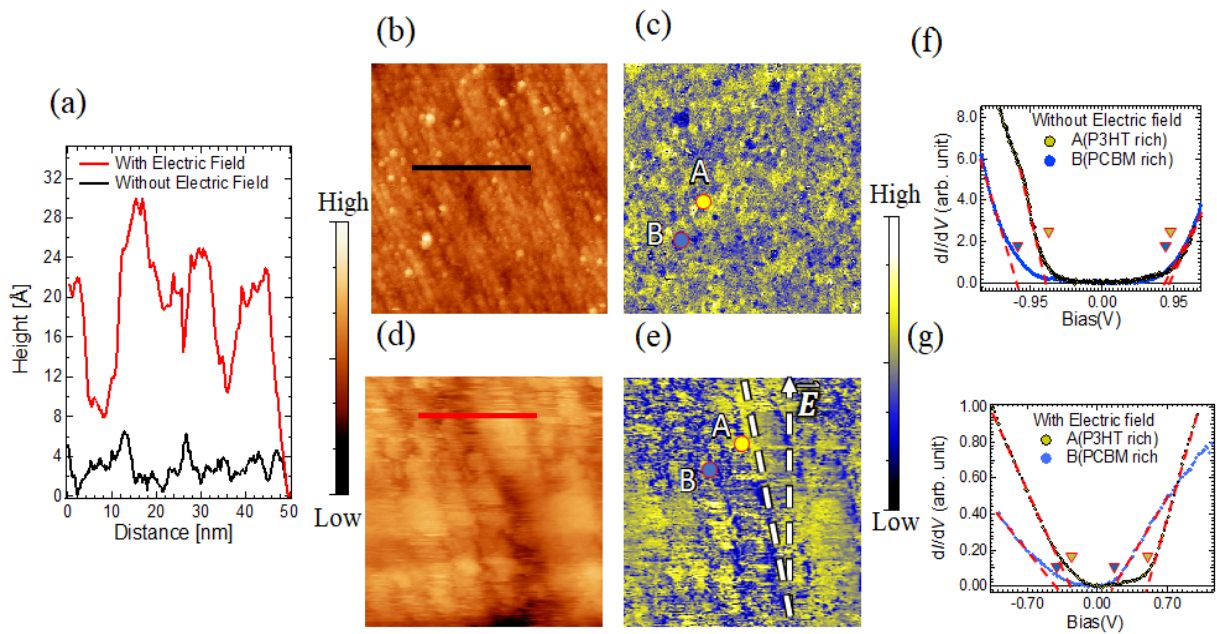

Figure 2. (**a**) Height profiles along the lines indicated in (b) and (d). STM topography images of samples (**b**) without electric field treatment and (**d**) with electric-field-assisted annealing treatment. Image size: 100 nm × 100 nm. d$I$/d$V$ mappings of samples (**c**) without electric field treatment and (**e**) with electric-field-assisted annealing treatment. Image size: 100 nm × 100 nm. In (**e**), the directions of the electric field and the elongated domain orientation are shown with white dashed arrow and line, respectively. d$I$/d$V$ point spectra measured on the samples (**f**) without electric field treatment and (**g**) with electric-field-assisted annealing treatment.

molecular domains and oriented near the electric field direction. The domain elongation is believed to be beneficial to the OPVC performance as the domains are better percolated toward the electrode directions [46]. This better percolated molecular domains might be the underlying reason of the measured improved charge mobility, hence the higher short circuit current ($J_{sc}$). Further analysis of the domain elongation will be discussed later in more details.

Here, another important observation can be made – intermolecular interactions induced doping effects. In particular, the d$I$/d$V$ point spectra taken at the high and low contrast regions in both samples are shown in Fig. 2(f) and (g). The HOMO and LUMO energies of each spectrum are labeled with inverse triangles. Based on the HOMO and LUMO energies, the P3HT-rich (donor) and PCBM-rich (acceptor) domains are assigned to the high and low contrast regions, respectively, in both samples. Surprisingly, the d$I$/d$V$ spectra and the energy gaps are different compared between the samples with and without the electric-field-assisted annealing treatment. The band gaps for the samples without electric field treatment are: 1.62 eV for the P3HT-rich regions and 1.95 eV for PCBM-rich regions. On the other hand, the band gaps for the samples with electric-field-assisted annealing treatment are: 0.77 eV for the P3HT-rich regions and 0.59 eV for PCBM-rich regions. The lower band gaps and the smaller HOMO/LUMO offset between the donor and acceptor regions in the electric field treated samples will lower the charge separation ability at the domain interfaces, and it may also lower the open circuit voltage ($V_{oc}$). It is known that the $V_{oc}$ of the OPVC is primarily related with the energy difference between the LUMO of the acceptor and the HOMO of the donor. In the data collected here, the $LUMO_{acc} - HOMO_{don}$ value changes from 1.57 eV (annealed) to 0.44 eV (electric-field-assisted annealed). It is believed here that the changes of the electronic properties in the molecular domains in the two samples are due to a different level of PCBM-P3HT intermixing, which cause intermolecular interaction induced doping effects in the P3HT-rich or PCBM-rich domains [34]. This hypothesis is further supported by the XRD data shown later. In literature, this lowering $V_{oc}$ upon the electric-field treated was also reported in some systems [17,27]. Thus, the overall OPVC performance in the electric-field-assisted annealed samples is the result of the competing effects of elongated domains (beneficial effects) and the intermolecular interaction induced doping (negative effects).

The elongated domains seen in d$I$/d$V$ mapping are further quantified by analyzing the two-dimensional (2D) Fourier Transform (FT) images of the d$I$/d$V$ mappings. Figure 3(a) and (c) shows the FT images of the d$I$/d$V$ mappings shown in Fig. 2(c) and (e), respectively. It is already obvious that the FT images exhibit isotropic pattern (round shape feature near $\vec{k} = 0$) for the samples without electric field treatment and anisotropic pattern (ellipse shape feature near $\vec{k} = 0$) for the samples with electric-field-assisted annealing treatment , as shown in the zoom-in FT images in Fig. 3(b) and (d). The round/ellipse features near $\vec{k} = 0$ were fit by a rotated ellipse Gaussian function as:

$$F(k_x, k_y) = Ae^{-(\frac{((k_x-k_{x_0})sin(\theta) + (k_y-k_{y_0})cos(\theta)))^2}{a^2} + \frac{((k_x-k_{x_0})cos(\theta) - (k_y-k_{y_0})sin(\theta))^2}{b^2})} \qquad (1)$$

where *A* is the intensity; $\theta$ is the angle between the electric field and minor axis in reciprocal space which is directly related to the orientation of the domains in real space as shown in Figure 3(d); *a* and $b$ are the lengths of the major and minor axes, respectively; while $k_{x_0}$ and $k_{y_0}$ are the coordinates of the center. Figure 3 (b) and (d) show the zoom-in view of the FT images of the d$I$/d$V$ mappings near $\vec{k} = 0$ and the fitting results (contour plots). Two main quantities are extracted: (1) the eccentricity ($\epsilon = \sqrt{1 - (\frac{b}{a})^2}$, which ranges from 0 for perfect circle to 1 for extreme ellipse), and (2) the orientation angle $\theta$. From Fig. 3(b), the eccentricity is determined to be $0.1 \pm 0.1$, indicating the isotropic texture. On the other hand, for Fig. 3(d), the eccentricity is determined to be $0.84 \pm 0.01$, indicating anisotropic texture. The orientation of the elongated domains with the electric-field-assisted annealing treatment is found to be $11.5° \pm 1.5°$. Based on the measured six independent d$I$/d$V$ mappings, the statistics of the eccentricities in the samples without the electric field treatment were determined to be $0.13 \pm 0.06$ (see details in Supporting Information, Table S1). On the other hand, same analysis was performed on six independently measured d$I$/d$V$ mappings of the samples with electric-field-assisted annealing treatment. The statistics of the eccentricity and the angle of the elongated domains are determined to be $0.824 \pm 0.004$, and $6.2° \pm 4.8°$, respectively (see details in Supporting Information, Table S2).

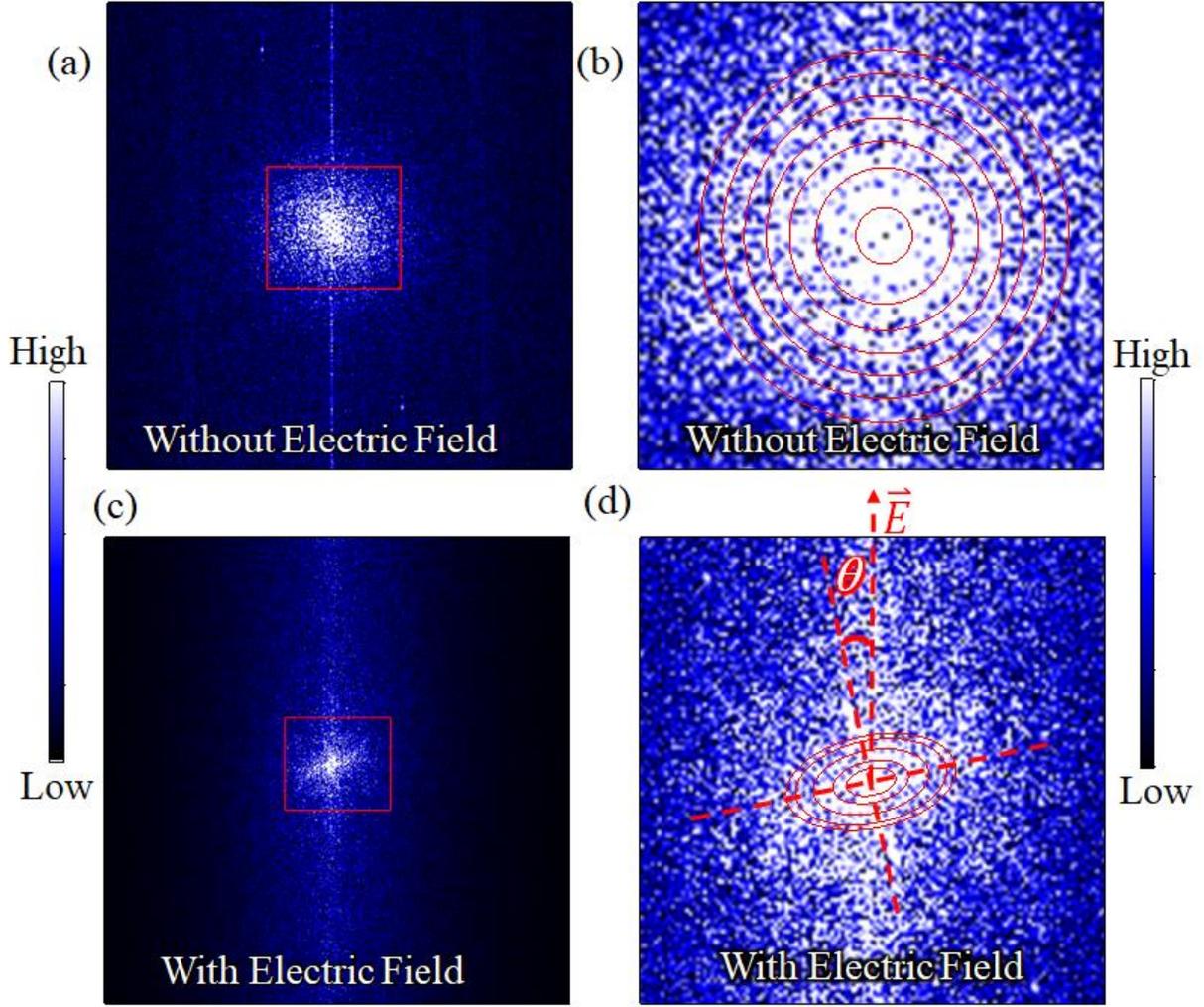

Figure 3. Fourier transform images of d*I*/d*V* mappings (Image size: 2.20 [1/nm] × 2.20 [1/nm]) for samples (**a**) without electric field treatment and corresponding zoom-in views (Image size: 0.73 [1/nm] × 0.73 [1/nm]): (**b**) and (**d**).

The molecular intermixing and the domain-domain separation are also confirmed by the XRD and SAXS measurements. Figure 4(a) shows the XRD data (substrate signals were removed) collected on samples with conditions of (i) unannealed; (ii) annealed; and (iii) electric-field-assisted annealed. The P3HT (100) diffraction peak is clearly observed in all three samples. The peaks were fit with a Gaussian equation to extract the peak positions and widths, as summarized in Table S3 in Supporting Information. The peak positions are found to be shifted from the unannealed $5.30° \pm 0.04°$ value to $5.24° \pm 0.03°$ upon annealing and to $5.18° \pm 0.04°$ for the electric-field-assisted annealed samples. It was reported that for the pure P3HT, the (100) peak position is in the range of 5.30° - 5.59° [42,47–50]. The down shifting in the XRD peak position in the samples reported here compared to the pure P3HT case indicates that the P3HT lattice periodicity increased with the incorporation with the PCBM molecules. The evolution of the $2\theta$ angles from the unannealed, to annealed and finally to the electric-field-assisted annealed samples indicate that the P3HT/PCBM intermixing is getting more severe. Similar trend was also observed in previous reported XRD data [15,16]. The peak width was reduced from 0.60° ± 0.06° to 0.44° ± 0.05° upon annealing and remain the same (0.42° ± 0.06°) with the electric-field-assisted annealing, indicating a better crystallinity is achieved by the annealing treatment. It thus can be concluded here that the annealing process makes a better crystalline P3HT domains and the PCBM molecules are incorporated to expand the periodicity with the annealing and further promoted by the electric-field-assisted annealing process. This result also pointed out that the electric field treatment may have the ability to control the level of the P3HT-PCBM intermixing. This conclusion also strongly supports the intermolecular interaction induced doping effects found in the d$I$/d$V$ spectra (Fig. 2(f) and (g)).

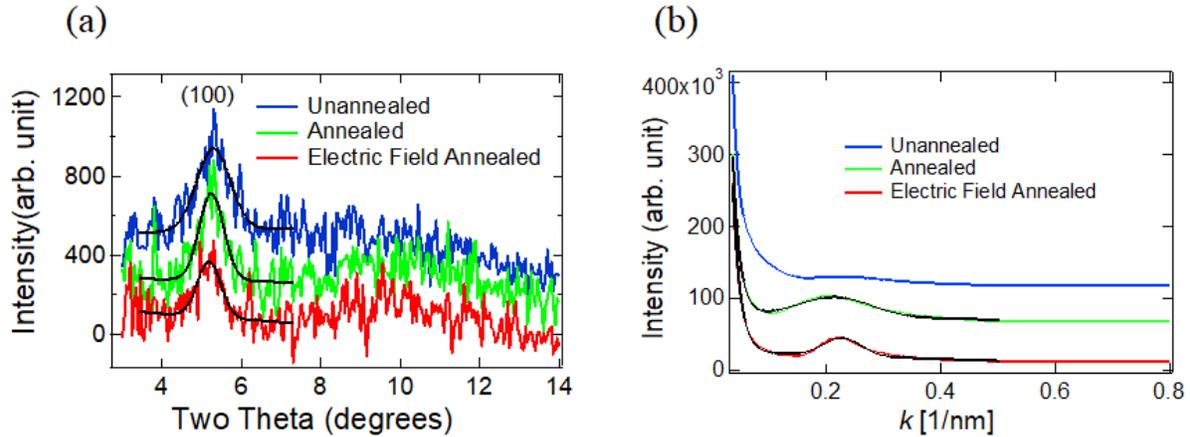

Figure 4. (a) XRD data and (b) SAXS data for three samples, unannealed, annealed, and electric-field-assisted annealed.

While XRD provides the molecule-molecule crystallinity information, the larger scale domain-domain correlation information can be extracted from SAXS measurements. Figure 4(b) shows SAXS data measured on the above mentioned three samples. The SAXS data of the P3HT/PCBM mixture can be described with a decay curve plus a peak feature [40,51,52]. It was pointed out that the SAXS peak position is closely related to the phase separation length scale (domain-domain distance) or also known as the domain correlation length [51–54]. The peak width is, then, related to the uniformity of the domain-domain distance. The SAXS data was fitted with an exponential decay plus a Gaussian function to extract the peak information, which are summarized in Table S4 in Supporting Information. Note that for the unannealed samples, no fitting was performed as there is no clear peak (only an unclear hump was observed). The annealed samples show a wider peak ($0.089 \pm 0.004$ nm$^{-1}$) compared to that of the electric-field-assisted annealed samples ($0.046 \pm 0.002$ nm$^{-1}$), indicating that the electric-field-assisted annealed samples have a more uniform domain-domain distance. This agrees well qualitatively with the observed d$I$/d$V$ mapping (Fig. 2(c) and (e)), where Fig. 2(c) shows various sizes of domains (hence the wider range of domain-domain distances) and Fig. 2(e) has more uniform domain sizes (thus more uniform domain-domain distance). The peak positions of the two samples, annealed and electric-field-assisted annealed, are found to be 0.216 nm$^{-1}$ and 0.228 nm$^{-1}$, respectively. They are corresponding to the domain-domain distances of 29.1 nm and 27.6 nm, respectively. Again, this observation is qualitatively agreeing with the d$I$/d$V$ mappings (Fig. 2(c) and (e)).

**Conclusion**

In short, with the data presented here, it is concluded that the electric-field-assisted annealing process should affect the molecular domain nanomorphology as well as the level of the molecular intermixing, which promotes the intermolecular interaction induced doping effects.

Elongated domains are observed to have the domain eccentricity of $0.824 \pm 0.004$ and orientation aligned along with the electric field with an angle of $6.2° \pm 4.8°$. The elongated domains with the application of electric field are beneficial for the charge transport, hence better $J_{sc}$ for the OPVC performance. On the other hand, the d$I$/d$V$ point spectra (Fig. 2(g)) for the electric-field-assisted annealed samples show smaller band gaps and smaller $LUMO_{acc} - HOMO_{don}$ values. This altered relationship of the energy levels has a negative impact on the $V_{oc}$ for the OPVC performance. The moderate performance improvement [15,16,18,21,22] upon the electric-field-assisted annealing treatment is a result of these two competing factors. This work points out that in additional to the nanomorphology control, the molecular intermixing may be equally important in impacting the OPVC performance.

**Acknowledgement**

Authors acknowledge the grant support from the National Science Foundation (NSF) under Award Nos. CMMI-1662435, 1662509 and 1753770.

**Author Contributions**

RD performed all the experiments. RD and TC analyzed the experimental results. RD and TC wrote the manuscript. All authors edited the manuscript and helped with the analysis of the results. WC, TC and GB conceived the project. The manuscript was written through contributions of all authors.